\documentclass{mn2e}
\usepackage{amssymb}
\input epsf.sty
\newif\ifAMStwofonts
\AMStwofontstrue

\def\fek{Fe K{\small $\alpha$}}
\def\Rblr{R_{\rm BLR}}
\def\fwhm{v_{\rm FWHM}}
\def\vr{v_{\rm r}}
\def\vp{v_{\rm p}}
\def\rms{$\sigma^2_{\rm nxs}$}
\def\mbh{$M_{\rm BH}$}
\def\mbhk{$M_{\mathrm{BH},k}$}
\def\mbhrev{M_{\rm BH, rev}}
\def\mbhdisp{M_{\rm BH, disp}}
\def\mbhvar{M_{\rm BH, \sigma^2}}
\def\msun{\rm M_{\odot}}
\def\tA{\widetilde{A}}

\def\f2{$f^2$}
\def\fsig{$f_{\sigma}$}
\def\four{1.12}

\def\fOnken{1.37}

\begin{document}

\title[Black hole mass determination in AGN]
{Consistency of the black hole mass determination in AGN
from the reverberation and the X-ray excess variance method}

\author[Niko\l ajuk, Czerny, Zi\' o\l kowski, Gierli\' nski]
 {M.~Niko\l ajuk$^1$, B.~Czerny$^2$, J.~Zi{\'o}\l kowski$^2$ and 
  M.~Gierli\' nski$^{3,4}$ \\
  $^1$Institute of Theoretical Physics, University of Bia\l ystok,
   Lipowa 41, 15-424 Bia\l ystok, Poland \\
  $^2$Copernicus Astronomical Center, Bartycka 18, 00-716 Warsaw,
  Poland \\
  $^3$Department of Physics, University of Durham, South Road, Durham
  DH1 3LE, UK \\
  $^4$Obserwatorium Astronomiczne Uniwersytetu Jagiello{\'n}skiego, 30-244
  Krak{\'o}w, Orla 171, Poland 
}

\maketitle
\begin{abstract}
Values of black hole masses are frequently determined with the help of
the reverberation method. This method requires a specific geometrical
factor related to the distribution of the orbits of the Broad Line
Region clouds. Onken et al. determined the value
\f2~$= 1.37 \pm 0.45$ from the black hole mass - dispersion relation.
In this paper we determine this factor using an independent mass
determination from the X-ray variance method for a number of Seyfert 1
galaxies and comparing them with the reverberation results by Peterson
et al. We obtain mean value \f2~$= 1.12\pm 0.54$, consistent
with Onken et al.  Both values are larger than the value 0.75
corresponding to a spherical geometry. It indicates that most probably
all values of the black hole masses obtained with the use of the Kaspi
et al. formulae should be multiplied by a factor of $\sim$ 1.7.
This also shows that the Broad Line Region is rather flat, and hints
for a dependence of the factor \f2 on a source inclination seem to be
present in the data.
\end{abstract}


\section{Introduction}

The value of the black hole mass is one of the key parameters in the
description of the accretion process in an active galactic nucleus
(AGN). Much effort was devoted to develop reliable and efficient
methods of mass determination from the observational data. Current
popular methods include the reverberation mapping (Blandford \& McKee
1982; Wandel, Peterson \& Malkan 1999; Kaspi et al. 2000; Onken \&
Peterson 2002; Vestergaard 2002), the stellar and gas kinematics
(Kormendy \& Richstone 1995; Nelson \& Whittle 1995; Ferrarese \&
Merritt 2000; Gebhardt et al. 2000; Woo \& Urry 2002; Tremaine et al
2002; Verolme et al. 2002; Peterson et al. 2004; Onken et al. 2004),
determination via water maser emission (Miyoshi et al. 1995; Greenhill
et al. 2003), the method based on BH mass -- bulge mass or bulge --
luminosity relation (Magorrian et al. 1998; Kormendy \& Gebhardt 2001;
McLure \& Dunlop 2002; H\" aring \& Rix 2004), based on the disk
luminosity estimation (Tripp, Bechtold \& Green 1994; Collin et
al. 2002) or just the $H_{\alpha}$ line analysis (Green \& Ho
2005). These methods were recently supplemented by the methods based
on X-ray variability: power spectrum break (Papadakis 2004; McHardy et
al. 2005) and excess variance (Nikolajuk et al. 2004; O'Neill et
al. 2005).

Each of the methods require the knowledge of a certain normalization
factor, and is biased by some systematic errors, so the comparison of
two independent methods is extremely important.

In the present paper we compare the results of the reverberation approach
with the results of the X-ray excess variance for several Seyfert 1 galaxies.
The methods of mass determination and the approach to their comparison is
given in Section~\ref{sect:method}, the results for mean normalization factors
and the dependence on the source inclination are shown in
Section~\ref{sect:results}, and we discuss the results in
Section~\ref{sect:discussion}.

\section{Method}
\label{sect:method}

\subsection{Selected sample of Seyfert galaxies}
\label{sect:sample}

We consider a sample of Seyfert 1 galaxies which were both a subject
of optical and X-ray monitoring (see Table~\ref{tab:sources}). We do
not include Narrow Line Seyfert 1 galaxies since there are many
indications of peculiar properties of these sources in comparison to
normal Seyfert 1 objects (smaller black hole mass to bulge mass ratio,
Wandel 1999; Mathur, Kuraszkiewicz \& Czerny 2001, higher variability
amplitude, Bian \& Zhao 2003; Nikolajuk, Papadakis \& Czerny 2004;
Markowitz \& Edelson 2004). In particular, we do not include NGC 4395
since it might be a NLS1 galaxy (FWHM of H$\beta$ line 1500 km
s$^{-1}$; Kraemer et al. 1999). This leaves us with 13 sources with
black hole masses in the range $\sim 10^7 - 3 \times 10^8 \msun$.

We use the reverberation results of Peterson et al. (2004) since they
are based on the most advanced analysis of the line profiles.
Most of the X-ray variability data were taken from \textit{RXTE} and
\textit{ASCA} databases (see Nikolajuk et al. 2004).  For F9 we take
the excess variances and the durations of all 7 observations from
Turner et al. (1999).  Turner et al.  give both the full duration of
the observation and the effective time actually covered by the
data. We use the full duration $T_{\rm D}$ as $T$ since gaps in the
data affect less the variance than the extension of the
monitoring. This is related to the steep power spectrum and the
dominance of the lower frequencies in variability.

Measurements of the inclination angles of the sources were taken
mostly from Nandra et al. (1997b). For IC 4329A, NGC 5548 and NGC 4593
we took inclinations from Mushotzky et al. (1995) and Guainazzi et
al. (1999), correspondingly.  These measurements are based on the
shape of the iron \fek\ line. For 3C 120 and 3C 390.3 we adopted the
values estimated by Eracleous \& Halpern (1998) and Ballantyne et
al. (2004) from observations of radio jets.

\subsection{Black hole mass from the reverberation method}

Reverberation method allowed to measure the mass directly in several
Seyfert galaxies (Wandel et al. 1999; Kaspi et al. 2000). The basic
formula underlying the reverberation method reads
\begin{equation}
M = f^2 ~ \frac{v_{\rm FWHM}^2 R_{\rm BLR}}{G} \ ,
\label{eq:rev_mass}
\end{equation}
where $M$ is the black hole mass, $G$ -- the gravitational constant,
$\Rblr$\ -- the radius of the Broad Line Region, $\fwhm$\ -- the Full
Width Half Maximum (FWHM) of $H\beta$ line and \f2\ is the squared
geometrical factor describing the distribution of the orbits of Broad
Line Region (hereafter BLR) clouds. In a recent more sophisticated
approach of Peterson et al. (2004) to the line profile analysis the
$\fwhm$ was replaced with the dispersion $\sigma_{\rm line}$:
\begin{equation}
M = f_{\sigma} ~ \frac{\sigma^2_{\rm line} \, c \tau_{\rm cent}}{G} \ ,
\label{eq:rev_mass2}
\end{equation}
where $ \tau_{\rm cent}$ is the centroid of the cross-correlation
function, $c$ is the light velocity.
In the sample of Peterson et al. statistically there is a relation
between the two factors
\begin{equation}
f^2 \approx \frac{1}{4} f_{\sigma} \ .
\label{eq:relation}
\end{equation}

Onken et al. (2004) determined the scaling coefficient in equation
~(\ref{eq:rev_mass2}) from the correlation between the black hole mass
and the bulge/spheroid stellar velocity dispersion. Their value of
\fsig$= 5.5 \pm 1.8$ was adopted by Peterson et al. (2004). We use
their results in our inclination-independent approach. In this
approach there is no space for any anisotropy of the BLR.

However, BLR is unlikely to be spherically symmetric (e.g. Done
\& Krolik 1996; Krolik 2001, Collin \& Kawaguchi 2004).
Most plausible geometry, particularly for the Low Ionization Line
Region, is a disk-like wind (Collin-Souffrin et al. 1986; Chiang \&
Murray 1996; Hutchings et al. 2001; Kollatschny 2003). Unfortunately,
the theoretical predictions of the shape of an emission line is quite
sensitive to the assumptions about the optical depth of the wind and
the distribution of the emissivity (Murray \& Chiang 1997).

Therefore, we consider, as a possibility, that the BLR velocity may be
represented by a combination of a random isotropic component, $\vr$,
and a component only in plane of the disc, $\vp$ (Wills \& Browne
1986; McLure \& Dunlop 2001). Therefore, the masses of the black hole
in AGNs can be represented by
\begin{equation}
M = \frac{1}{4(\xi^2 +  \sin^2 i)} \frac{\fwhm^2 \Rblr}{G} \ ,
\label{eq:mbhsin1}
\end{equation}
where $\xi = \vr/\vp$. It differs by a factor of two in front of $\sin i$
from the formula of Krolik (2001) and of Collin \& Kawaguchi (2004).

This relation means that \fsig\ is not an inclination-independent
coefficient but
\begin{equation}
f_{\sigma} = \frac{1}{\xi^2 + \sin^2 i} \ .
\label{eq:mbhsin2}
\end{equation}
If $\xi = 0$, the value \fsig$=5.5$ determined by Onken et al. (2004)
corresponds to a representative inclination angle $i = 25.2^{\circ}$.

The use of this formula requires the knowledge of $\xi$ and $i$ for
every object separately, and the exact dependence on the inclination
angle is not well justified. However, it is useful for search of any
possible traces of anisotropy.

\subsection{Black hole mass from the X-ray variability}
\label{sect:varmeth}

\begin{table*}
\caption{The sources used in our study.
}

\label{tab:sources}
\begin{center}
\begin{tabular}[t]{l l c c c c c c c c c c}
\hline\hline
Name  & Type   & $\mbhvar^{\rm Peters}$
& $\mbhvar$ &  $r=\frac{\mbhvar}{M_{\rm BH, rev}^{\rm Peters}}$ &
$\log L_X$ & $\Gamma_{3-10}$ & $W_{\rm K\alpha}$ & $i_{\rm Nandra}$
& $i_{\rm  oth.papers}$(ref.) & $i$ \\

      &        & ($10^{7}\ \rm M_{\odot}$) & ($10^{7}\ \rm M_{\odot}$)
& & &         &   (eV)  & (deg) & (deg) & (deg) & \\
\hline
3C 120    & S1.5  & $5.55^{+3.14}_{-2.25}$  & ${19.7}^{+20.4}_{-8.9}$
& $3.55 \pm 3.11$ & 43.95$^{\dag}$ & 1.79 & 70 &
$60^{+30}_{-14}$ & $<14^{(b)}$(1,2) & $13^{+6}_{-6}$ \\

3C 390.3  & S1.5  & $28.7 \pm 6.4$ & $22.3^{+14.8}_{-8.5}$
&$0.78 \pm 0.44$  & 44.20$^{\dag}$ & &
& & $19\leq i \leq 33^{(b)}$(2) & $29^{+10}_{-9}$ \\

Ark 120   & S1.0  & $15.0 \pm 1.9$  & $10.8^{+16.3}_{-6.3}$
& $0.72 \pm 0.72$ & 43.88 &  &  &  &  & $30^{+20}_{-17}$ \\

IC 4329A  & S1.2  &$< 2.78$ & $5.42^{+3.62}_{-1.96}$
& $>1.95$ & 43.59 & 1.70 & 80 & $22^{+15}_{-22}$ &
$<24^{(a)}$(3) & $<18$ \\

Mrk 509   & S1.5  & $14.3 \pm 1.2$  & $8.00^{+4.00}_{-2.41}$
& $0.56 \pm 0.23$ & 44.03 & 1.76 & 70 &
$41^{+49}_{-32}$ &   & $35^{+9}_{-8}$ \\

NGC 3227  & S1.5  & $4.22 \pm 2.14$  & $4.11^{+1.76}_{-1.17}$
& $0.97 \pm 0.60$ & 41.66 & 1.52 & 120 &
$20^{+10}_{-10}$ & & $26^{+9}_{-8}$ \\

NGC 3516  & S1.5  & $4.27 \pm 1.46$  &  $2.95^{+1.60}_{-0.85}$
& $0.69 \pm 0.37$ & 43.08 & 1.74 & 90 &
$27^{+4}_{-5}$  & & $31^{+10}_{-9}$ \\

NGC 3783  & S1.5  & $2.98 \pm 0.54$  &  $2.13^{+0.92}_{-0.70}$
& $0.71 \pm 0.30$ & 42.90 & 1.54 & 115 & $26^{+9}_{-10}$ & &
$30^{+7}_{-7}$ \\

NGC 4151  & S1.5  & $1.33 \pm 0.46$  &  $2.92^{+0.97}_{-0.70}$
& $2.19 \pm 0.98$ & 42.62 & 1.55 & 100 & $23^{+12}_{-13}$ &
& $17^{+4}_{-4}$ \\

NGC 4593  & S1.0  & $<1.47$ & $0.67^{+0.29}_{-0.45}$ &
$>0.45$           & 42.98 & 1.78 & 90 & $45^{+45}_{-45}$ &
$32^{+23 (a)}_{-12}$(4) & $<39$\\

NGC 5548  & S1.5  & $6.71 \pm 0.26$  & $14.1^{+21.1}_{-4.1}$
& $2.10 \pm 1.88$ & 43.41 & 1.75 & 90 & $47^{+43}_{-41}$ &  $15 \leq i \leq
38^{(a)}$(3) & $17^{+8}_{-8}$ \\

NGC 7469  & S1.5  & $1.22 \pm 0.14$  & $ 2.58^{+0.65}_{-0.51}$
& $2.11 \pm 0.53$ & 43.25 & 1.78 & 130 &
$19^{+71}_{-19}$ & & $17^{+2}_{-2}$ \\

F 9       & S1.2  & $25.5 \pm 5.6$  & $8.44^{+4.78}_{-2.24}$
& $0.33 \pm 0.15$ & 43.91& 1.83 & 120 &
$45^{+45}_{-21}$ & & $48^{+16}_{-12}$ \\
\hline\hline
\end{tabular}
\end{center}
Col.~(1) lists the object name and type (S1.0-S1.5 denote Seyferts
galaxies). Col.~(3) shows the masses of black holes obtained from
reverberation method by Peterson et al. (2004), who take the
geometrical factor $\langle f_{\sigma} \rangle = 5.5$. Col.~(4) shows
the black holes masses obtained from the variance method. The ratios
of the two masses are given in Col.~(5). Col.~(6) Log of 2-10 keV
luminosity in units of erg s$^{-1}$. All luminosities are given for:
$H_0 = 75$ km s$^{-1}$ and $q_0 = 0.5$. The values with the
superscript $^{\dag}$ were taken from Green et al. (1993) and
transformed to this cosmology. The other values were directly taken
from O'Neill et al. (2005). In Col.~(7) and (8) are power-law index
and equivalent width of Fe K$\alpha$ line, respectively. All values
were taken from Nandra et al. (1997b).  Col.~(9) shows the inclination
angles given by Nandra et al. (1997b). Col.~(10) shows the same
quantity but taken from other papers. The superscripts $^{(a)}$ and
$^{(b)}$ in Col. (10) denote inclination obtained from observation of
the \fek\ line and the radio jet correspondingly. The numbers in
parentheses correspond to the following references: (1) Ballantyne
et al. (2004), (2) Eracleous \& Halpern (1998), (3) Mushotzky et
al. (1995), (4) Guainazzi et al. (1999). Col.~(11) shows the
inclination angle calculated from equation~(\ref{eq:inc}) under the
assumptions $\xi = 0$.
\end{table*}

We determine the black hole mass using the excess variance method.
The method is based on assumptions that (i) the high frequency tail of
the power spectrum has a slope of -2 (ii) the high frequency break
scales inversely with the black hole mass (iii) the value of the power
times frequency at the high frequency break is universal for all
objects, independent from mass. In this paper we improve slightly this
method, in comparison to the method described by Nikolajuk et
al. (2004). We change: (i) the method of obtaining the final value of
\mbh\ for a given object from results for individual lightcurves (ii)
the method of calculation of the errors (iii) the value of the
normalization constant present in our method.

The X-ray variance method (Nikolajuk et al. 2004) is based on the
scaling of the normalized variance of X-ray light curves, \rms, with
black hole mass
\begin{equation}
M = C \frac{T - 2\Delta t}{\sigma^2_{\rm nxs}}.
\label{eq:var_mass}
\end{equation}
Here $T$ is the duration of a single X-ray light curve in seconds and
$\Delta t$ is its bin size in seconds, as well. \rms\ is in units of
(rms/mean)$^2$.  $T$ cannot be longer than the time scale defined by
the high frequency break of the power spectrum.  Longer light curves
can be chopped and used to determine several independent values of the
variance. The normalized excess variance, \rms, is defined as in
Nandra et al. (1997a) and Turner et al. (1999).

From each light curve of a given object we obtain the $k$th individual
black hole masses \mbhk , using equation~(\ref{eq:var_mass}). In order
to calculate the final value of \mbh ~ from several lightcurves, we
fit this set of individual \mbhk. For each lightcurve we calculate the
coefficient $\tA_k$
\begin{equation}
 \tA_k = {{(\sigma^2_{\rm nxs})}_k \over (T - 2 \Delta t)_k } ,
\end{equation}
and the final coefficient $\tA$ is calculated as a weighted mean from
the values of $\tA_k$, with the weight coefficient given by $[(T - 2
\Delta t)_k]^2$, i.e.
\begin{equation}
\tA = \frac{\sum_{k=1}^{N_{\rm lc}} (\sigma^2_{\rm nxs})_k (T - 2 \Delta t)_k}{\sum_{k=1}^{N_{\rm lc}} (T - 2 \Delta t)_k^2} \ .
\end{equation}
Here $(T - 2 \Delta t)_k$ and $(\sigma^2_{\rm nxs})_k$ are
results from several measurements. The number of the lightcurves is
$N_{\rm lc}$.

The final black hole mass in the object is derived directly from the
obtained value of $\tA$ and equation~(\ref{eq:var_mass})
\begin{equation}
M = \frac{C}{\tA} \ .
\label{eq:masa}
\end{equation}

The value of the constant $C$ in equation~(\ref{eq:var_mass}) must be
determined by applying the method to the source with a known mass. Cyg
X-1 was selected as a reference in the work of Nikolajuk et al. (2004).
The mass of the black hole in Cyg X-1 was assumed to be 10 M$_\odot$
and the value of the constant $C = 0.96 \pm 0.02$ M$_\odot$ s$^{-1}$
was derived. However, recent careful study of the evolutionary history
of this binary star showed that the most likely value for the mass of
the black hole in the system is $20 \pm 5$ M$_\odot$.  Therefore, we
assume 20 M$_{\odot}$ as the best estimate of the black hole mass in
Cyg X-1, and the constant $C$ in equation~(\ref{eq:var_mass}) is
consequently two times higher: $C = 1.92 \pm 0.5$ M$_\odot$ s$^{-1}$.
The error of this value is predominantly connected with uncertainty of
the Cyg X-1 black hole mass.

The accuracy of the black hole mass determination in the X-ray
variability method depends strongly on the effect of statistical error
of variance measurement and power leaking from long
timescales. Therefore, in our analysis we estimate the error by
performing Monte Carlo simulations for each source separately. A few
hundred sets of artificial data were generated for each of the
sources, using the Timmer \& K\"onig (1995) algorithm, exponent of the
lightcurve was calculated to account for the log-normal distribution
(see Uttley, McHardy \& Vaughan 2005), and each set was analyzed like
the data, thus giving a distribution of the values of $\tA$, so its
error was determined from the dispersion of the resulting
distribution. The mass error was determined from the error of $\tA$.

The error in the normalization constant was discussed separately since
the results of Peterson et al. (2004) also did not include the
coefficient error.

\subsection{Comparison of the two methods and search for inclination effects}
\label{sect:compa}

Assuming the absence of any inclination-dependent trend in BLR by
calculating the ratio $r$ of the two mass measurements for all objects
in our sample separately:
\begin{equation}
r = {M_{\rm BH,\sigma^2} \over M_{\rm BH,rev}^{\rm Peters}} \ ,
\end{equation}
where $M_{\rm BH,\sigma^2}$ is the black hole mass measured by the
excess variance method and $M_{\rm BH,rev}^{\rm Peters}$ is taken from
Peterson et al. (2004). We next calculate the mean and the median
values from the obtained distribution. If the obtained mean value
$\langle r \rangle$ is consistent with 1 within observational error,
the two methods statistically give consistent results.

Next we allow for an anisotropy of the BLR. We assume that the excess
variance method is independent on the source inclination but for the
reverberation results we adopt the coefficient \fsig\ given by
equation~(\ref{eq:mbhsin2}) instead of a fixed factor of 5.5. Under
this assumption the ratio $r$ of the two measurements should follow a
relation
\begin{equation}
r = {1 \over 5.5(\xi^2 + \sin^2 i)}.
\label{eq:ratsin}
\end{equation}
We check the possible presence of such trend using an independent
measurements of the sources inclinations (see
Section~\ref{sect:sample}).

We also show the results of a complementary approach, following the
general method of Wu \& Han (2001). We adopt the relation given by
equation~(\ref{eq:ratsin}), we determine the inclination angle of each
object from the formula
\begin{equation}
i = \arcsin \sqrt{\frac{1}{5.5 r} - \xi^2} \ ,
\label{eq:inc}
\end{equation}
and we compare them to independent measurements of the inclination.
Errors of the calculated inclination angle are based on the
error propagation theory (Bevington \& Robinson, 1969).

\section{Results}
\label{sect:results}

\subsection{Mass measurements}

\begin{figure}
\epsfxsize=8.8cm
\epsfbox[80 200 580 700]{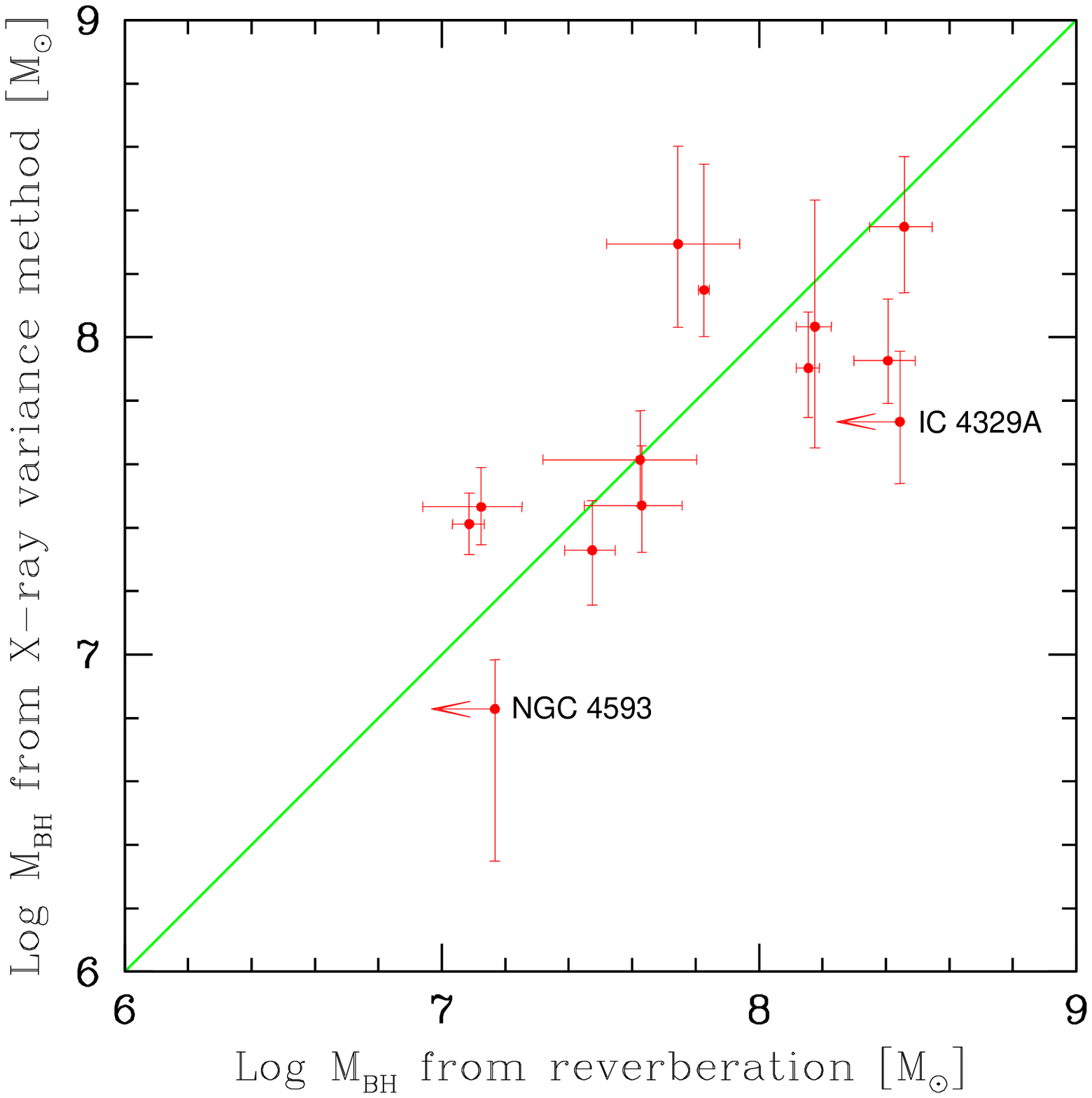}
\caption{Comparison of the black hole masses, $\mbhvar$,
taken from the X-ray variance method with the masses,
$\mbhrev$, taken from the reverberation method.
The solid line shows the expected $\mbhvar = \mbhrev$ relation.
}
\label{fig:MvarMrev}
\end{figure}

\begin{figure}
\epsfxsize=8.8cm
\epsfbox[80 200 580 700]{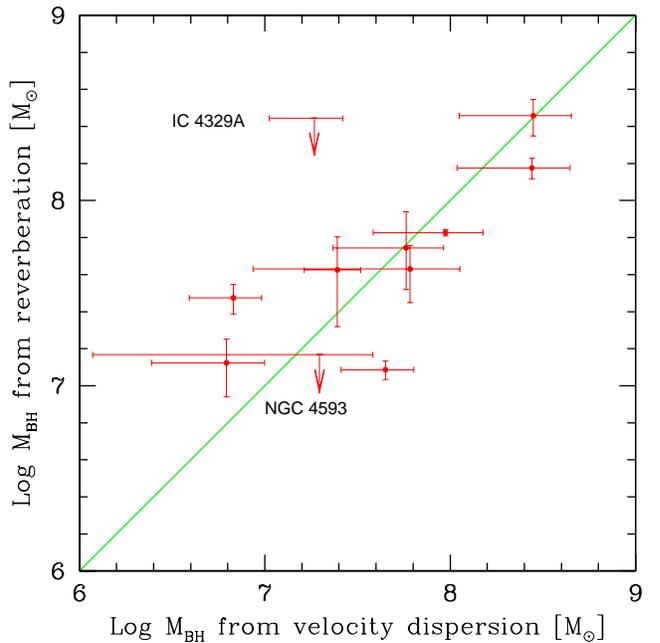}
\caption{The black hole masses obtained from the reverberation method
versus the masses of the same object obtained from the
\mbh--$\sigma_{\star}$ relation. The values of the $\mbhrev$ are taken
from Peterson et al. (2004).  The $\mbhdisp$ values of the velocity
dispersion method are calculated using $\sigma_{\star}$ values
obtained by Onken et al. (2004) and the formula of Tremaine et
al. (2002). The solid line shows the equality of the values $\mbhrev$
and $\mbhdisp$.  }
\label{fig:MrevMdisp}
\end{figure}

The values of the black hole masses, $M_{\rm BH,rev}^{\rm Peters}$,
determined by Peterson et al. (2004) with $\langle f_{\sigma}
\rangle = 5.5$ are given in the third column of Table~1. In fourth
column we give the masses $M_{\rm BH,\sigma^2}$ determined with the
excess variance method, as described in
Section~\ref{sect:varmeth}. Generally, the two measurements correlate
quite nicely, as shown in Fig.~\ref{fig:MvarMrev} and 
Fig.~\ref{fig:MrevMdisp}.

Statistical measurement errors are usually smaller in the case of
reverberation method. On the other hand, in two cases (IC 4329A and
NGC 4593) reverberation method gave only upper limits while the excess
variance method gave both the upper and lower limits, consistent
within two sigma error with the results from the velocity dispersion
method.

In fifth column of Table~1 we give the mass ratio $r = {M_{\rm
BH,\sigma^2}/M_{\rm BH,rev}^{\rm Peters}}$.  The distribution shows
significant scatter, with the minimum value obtained for Fairall~9 and
the maximum value obtained for 3C 120. Therefore, we calculate the mean
and the median value. In calculations of the weighted mean we exclude
IC~4329A and NGC~4593, because of the upper limits in the determination
of $\mbhrev$ from those galaxies. The weighted mean is equal to $0.81
\pm 0.15$. The given error is only statistical one sigma error. This
indicates that both methods are normalized rather accurately and there
is no strong need to introduce the systematic errors due to \fsig\ and
$C$ constants.  This is very encouraging since the level of
disagreement is much lower than allowed for by systematic errors
entering the two methods as errors of the normalization constants:
\fsig$ = 5.5(1 \pm 0.33)$ and $C=1.9 (1 \pm 0.26)$  M$_\odot$ s$^{-1}$.
It also means that indeed statistical errors are likely to be
dominating individual measurements in both methods.

No apparent trend with the black hole mass is present, so probably no
large systematic error related to the black hole mass is involved.

\begin{figure}
\epsfxsize=8.8cm
\epsfbox[80 200 580 700]{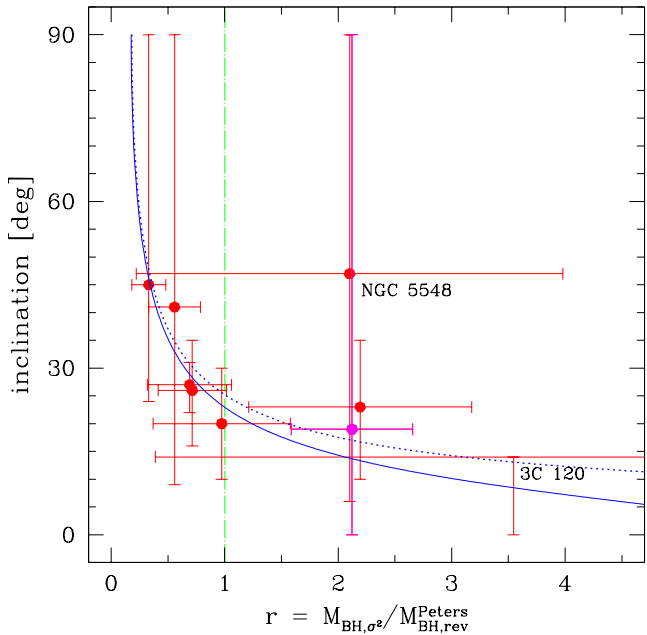}
\caption{Mass ratio $r$ versus inclination taken from literature
($i_{\rm Nandra}$ and $i_{\rm oth.papers}$; see references in Table~1).
Continuous line shows the best fit to the data under assumption that
$\xi = 0.19$. Dotted line represents the same fit but for $\xi = 0.0$
(see text for details).
}
\label{fig:inclicor}
\end{figure}

\subsection{Search for BLR anisotropy}

If the BLR is anisotropic this anisotropy is likely to show up in a
different way in the two black hole mass measurement
methods. Therefore, in Fig.~\ref{fig:inclicor} we show the relation
between the mass ratio, $r$, and the inclination angles measured
independently (see Section~\ref{sect:sample}). Errors are large but
the two quantities seem to be correlated: larger mass measurement
ratios are more likely to be derived for lower source inclination. The
Spearman rank correlation test for all ten pairs does not confirm this
impression ($R = 0.633$). However, if the inclination angle of NGC
5548 (47$^{\circ}$) from Nandra et al. (1997b) is replaced with the
value of 26.5$^{\circ}$ (mean value from Mushotzky et al. 1995, see
column (10) of Table~1, also from the iron line profile), the presence
of the correlation is confirmed ($R = 0.85$). Without this object, the
correlation is excellent ($R = 0.975$).

In order to investigate an importance of our correlation we check
up if the mass ratio correlated with power law index, $\Gamma_{\rm
3-10\ keV}$, or with line equivalent width, $W_{\rm K \alpha}$.
We take the appropriate values in Nandra et al. (1997b), and
we include them in Table~\ref{tab:sources}.
If we do not find any correlation between $\Gamma_{\rm 3-10\ keV}$ and
$r$ and/or between $W_{\rm K \alpha}$ and $r$ then our correlation,
the mass ratio-inclination angles, is more believable.
For the sample of our objects we do the Spearman rank
correlation test and in both cases we do not confirm any correlation.
In the case of $r$-$\Gamma_{\rm 3-10\ keV}$ a Spearman Correlation
Coefficient $R = -0.06$. In the second case, where we test $r$-$W_{\rm
K \alpha}$ once again the coefficient $R$ does not show any correlation
($R = -0.07$).

Uttley and McHardy (2005), McHardy et al. (2005) have suggested that
the deviation from linear scaling between the high frequency break and
$M_{\rm BH}$ may by a function of accretion rate.  In order to check
this hypothesis and its possible influence to our mass estimation,
$\mbhvar$, we examine a new correlation.  We check whether $r$
correlate with the ratio $L_X/\mbhvar$ (i.e. the ratio of 2-10 keV
luminosities divided by $\mbhvar$). We use sample of eleven objects
(all those objects, which have calculated $r$) and find an appropriate
values of $L_X$ in papers of O'Neill et al. (2005) and Green et
al. (1993).  The cosmological parameters which we use are: $H_0 = 75$
km s$^{-1}$ and $q_0 = 0.5$. The Spearman Correlation Coefficient $R$,
which we obtained, is equal to -0.54. We can say that $r$ probably
slightly anticorrelates with the ratio $L_X/\mbhvar$. Nevertheless, it
is difficult clearly confirm this hypothesis, because only $|R|> 0.73$
(for the sample of 11th pairs) indicates the presence of the correlation
at the 95\% level of confidence.

Assuming a BLR model
outlined in Section~\ref{sect:compa} we expect the inclinations to
follow the equation~(\ref{eq:inc}). Since $\xi^2$ in
equation~(\ref{eq:inc}) is unknown, we attempted to determine it from
the requirement of the best fit to the data. We searched for a minimum
of the function:
\begin{equation}
\chi^2 = \sum_{n=1}^{10} { (i_{obs}-i_{anal})^2\over (\delta i)^2},
\end{equation}
where $(\delta i)^2$ included both the measurement error of the
inclination angle (given in Table~1) and the error in prediction of
$i_{anal}$ from equation~(\ref{eq:inc}) due to an error in the
measurement of $r$. Resulting $\chi^2$ is lowest for $\xi = 0.19$ but
the errors are so large that $\xi = 0.0$ is also acceptable. We show
both fits to the data in Fig.~\ref{fig:inclicor}.

One object (3C 120) shows significant departure from the overall
mean. High value of the $r$ is likely to be due to an extreme value of
the viewing angle. Small inclination angle $i$ decreases the line
widths (see equation~\ref{eq:mbhsin2}).
3C 120 is most probably viewed face on since it shows superluminal
motion (e.g. Homan et al. 2001).

\begin{figure}
\epsfxsize=8.8cm
\epsfbox[80 200 580 700]{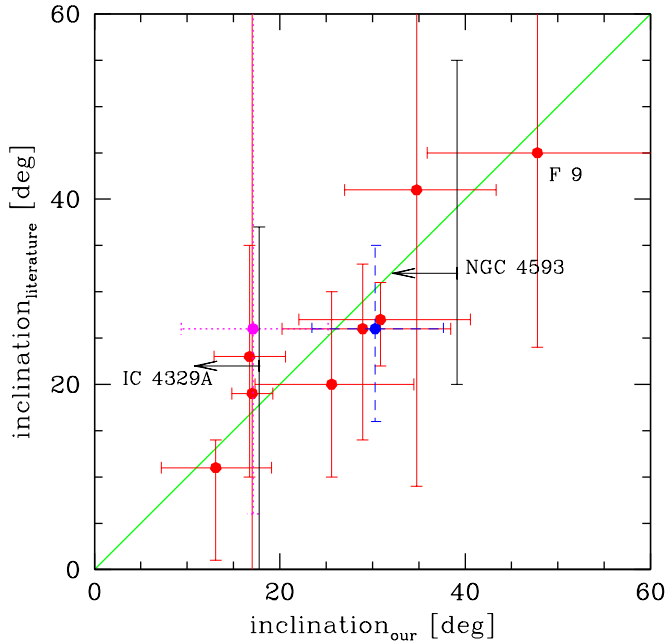}
\caption{Comparison of the inclination angles of the accretion discs
obtained from the X-ray variance method with the inclination angles
taken from different papers. The inclinations from literature are
derived mainly from \fek\ observations. Other values are based on the
radio jet observations. The references of the papers are included in
Col. (10) of the Table~1. The solid line shows the $i_{\rm
from~literature} = i_{\rm from~the~X-ray~variance}$ relation.
Different lines, which indicate errors of $i$ are only to avoid
mistakes and 'well' viewing. }
\label{fig:inclin}
\end{figure}

In Fig.~\ref{fig:inclin} we compare the determination of the
inclination from equation~(\ref{eq:inc}) (Table~1, last column) with
the results (mostly) from the shape of the iron line (see
Section~\ref{sect:sample}).  The agreement between the two methods is
encouraging. It both supports the existence of the anisotropic part in
the velocity field of BLR and the correctness of the inclination
determination by Nandra et al. (1997b). Recent results from XMM-Newton
telescope frequently do not support the existence of the broad
relativistically smeared iron line (e.g. Mkn 509, Page et al. 2003;
NGC 3227, Gondoin et al. 2003; NGC 4151, Schurch et al. 2003; NGC
5548, Pounds et al. 2003; NGC 3783, Reeves et al. 2004; NGC 4593,
Reynolds et al. 2004; NGC 3516, Turner et al. 2005). It is possible
that high quality data is sensitive to the departures of the actual
line from the fitted shape and reject the presence of such
feature. This departure may be, for example, related to the warm
absorber effect since narrow absorption features superimposed onto a
broad line modify the shape of the line making it more symmetric,
almost two-Gaussian (R\'o\. za\' nska et al. 2006).

We were also able to determine the inclination for Ark 120 (no
previous inclination measurements were done for this object). Measured
value is quite reasonable for a Seyfert 1 galaxy. It will be
interesting to see whether future independent determinations will
confirm our result.

\section{Discussion}
\label{sect:discussion}

\subsection{Consistency of black hole mass measurement methods}

We compare two methods of black hole mass determination:
reverberation method and X-ray excess variance method.
Both methods include important scaling factors.

Reverberation method is based on determination of the width of the
broad lines and the delay in their response to variable continuum. It
requires an assumption about the BLR geometry, in a form of a specific
value of the geometrical factor. Onken et al. (2004) obtained this
factor by using the black hole mass measurements derived from the
velocity dispersion measurements. This factor was used by Peterson et
al. (2004) for the reverberation measurements used in our paper.

Excess variance method is based on a scaling of the X-ray power
spectra with black hole mass. It requires an a~priori knowledge of
black hole mass in a reference object. It is convenient to use the
Galactic X-ray source Cyg X-1 in its hard state to scale the
method. The best current determination of the black hole mass in this
object is $20\pm 5 \msun$ (Zi\' o\l kowski 2005), and we adopt
this value in the present analysis.

We also attempted to use another Galactic source, GX~339-4, as a
reference. We have extracted power spectra from the hard state during
the rise of the 2002/2003 outburst (observations 0--24 from Belloni et
al. 2005) from the PCA onboard {\it RXTE}. We have corrected the
spectra for the Poisson noise and dead time effects of the instrument.
Then, we integrated them between 10 and 128 Hz in order to determine
the constant $C$. We have assumed the mass of 10 M$_\odot$ (Zdziarski
et al. 2004) and found $C_{\rm GX339-4} = (0.91 \pm 0.04)$ M$_\odot$
s$^{-1}$, by averaging over all observations. It is interesting to note
that the value of $C$ for individual observations remained roughly
constant throughout the wide range of observed luminosities (count rate
$\sim$400--900 $s^{-1}$ per one PCA detector). The resulting constant
$C$ is less than the value of $C_{\rm Cyg~X-1} = (1.92\pm0.5)$
M$_\odot$ s$^{-1}$, found in Section \ref{sect:varmeth}, but
uncertainties due to mass estimation are large. In particular, the mass
of GX 339--4 is not known very well, with only the lower limit from the
mass function, $5.8 \pm 0.5 \msun$ (Hynes et al. 2003), firmly
established. Higher mass would give higher $C$. We conclude that the
value of constant $C$ if likely to be between 1 and 2, though finding
its more exact value would require thorough investigation of many X-ray
binary sources. We plan to do this in the forthcoming paper.

We showed that the two methods of black hole mass determination, with
the scaling adopted as described, are statistically in agreement. The
mean value of the mass ratios is $r = 0.81 \pm 0.15$ (one sigma
error), consistent with 1. 
Recent determination of the black hole mass in NGC~4151 ($4.14 \pm
0.73 \times 10^7$ M$_{\odot}$) by Metzroth, Onken, Peterson (2006),
based on reverberation, agrees even better with our values (based on
excess variance) than the value given by Peterson et al. (2004) quoted
in Table~1.
This means that normalization factors used
by the two methods were determined correctly, and their corresponding
errors are perhaps even smaller than the conservatively estimated
error. Therefore, black hole masses can be measured rather accurately,
using any of the two methods if the requested data for an object is of
appropriate quality.

\subsection{Mean geometrical factor}

Since more discussions in the literature usually concerned the squared
geometrical factor \f2, as introduced in equation~(\ref{eq:rev_mass}),
we convert for convenience the discussed results \f2. The
normalization obtained by Onken et al. (2004) corresponds to \f2$=1.37
\pm 0.45$. Our results from the X-ray excess variance method gave
slightly lower values of the black hole masses so they effectively
correspond to \f2$=1.03$ (median value) and \four\ $\pm$ 0.20
(weighted mean). However, recently Collin et al. (2006) have obtained
\f2$=0.96$ from the reverberation method.

Kaspi et al. (2000) use equation~(\ref{eq:rev_mass}) and adopt the
value $f^2 = 3/4$ corresponding to a isotropic distribution of
randomly oriented orbits (Netzer 1990). They also show that a
convenient scaling exists between the continuum luminosity at $\lambda
= 5100$~\AA\ which is produced in accretion disc and the size of the
BLR, $\Rblr$. This scaling was used later in many papers (e.g. Woo \&
Urry 2002; Willot, McLure \& Jarvis 2003, Wu et al. 2004, Warner,
Hamann \& Dietrich 2004, Czerny, R\' o\. za\' nska \& Kuraszkiewicz
2004) for estimation of the black hole mass without a need for
time-consuming monitoring. Again, the factor $f^2 = 3/4$ was
assumed.

Results based on the excess variance support the conclusion of
Onken et al. (2004) that $f^2 = 0.75$ is certainly too low, and
the appropriate value is by a factor $\sim 1.7 - 1.8$ higher than
that.

This will increase black hole masses derived with the use of Kaspi et
al. (2000) formulae, and in consequence it will decrease the estimated
value of the Eddington ratio. It will (at least partially) reduce the
problem of the highly super-Eddington accretion rates in some quasars
(Collin et al. 2002).

Normalization higher by a factor of 2 than in the original formula of
Kaspi et al. (2000) also leeds to estimates of the quasar radiation
efficiency in better agreement with the timescales of the black hole
growth (Wang et at. 2006).

\subsection{Broad Line Region Geometry}

Attempts of the direct determination of the BLR geometry from the
reverberation studies of the line profiles in NGC 5548 (Done \& Krolik
1996) showed that none of the simple models reproduced the velocity
field of the clouds. It is possible that the Keplerian motion
dominates but the disk clumpiness is responsible for the observed
complexity of the picture (Shapovalova et al. 2004).

Our weighted mean value of \f2\ thus can be inverted into a mean value
of the inclination angle, $i$, as done by Wu \& Han (2001).  If we
neglect the random/wind component of the velocity (i.e. for $\xi= 0$),
mean value of \f2\ corresponds to the inclination angle $28 \pm 4
^{\circ}$, and the median value of \f2\ is equivalent to $i \simeq
23^{\circ}$.  Both values are reasonable, taking into account that
large viewing angles are obscured by the dusty/molecular torus and
active nuclei with large inclinations are classified as type 2
objects.

\section{Conclusions}

\begin{itemize}

\item We determine the squared geometrical factor $f^2$ present in the
reverberation method using black hole masses determined from X-ray
variance method as a reference.  Our value of the geometrical factor,
$f^2 =$ \four, is consistent with the value \fOnken\ obtained by
Onken et al. (2004) from the black hole mass - stellar dispersion
relation, and applied by Peterson et al.  (2004), and in agreement
with $f^2 = 0.96$ of Colin et al. (2006).  Such values are
intermediate between the expected values for a spherical and a flat
geometry of the BLR.

\item The values of the black hole masses given by the formulae of Kaspi
et al. (2000) are most probably systematically underestimated by a factor
of $\sim$ 1.7-1.8.

\item complementary use of the reverberation and excess variance
methods is a promising tool to constrain the inclination of an AGN

\end{itemize}

\section*{Acknowledgments}

We would like to thank Andrzej So\l{}tan for helpful comments and
useful discussions.  Part of this work was supported by grants
1P03D00829 and PBZ-KBN-054/P03/2001 of the Polish State Committee for
Scientific Research.

\ \\
This paper has been processed by the authors using the Blackwell
Scientific Publications \LaTeX\ style file.

\end{document}